\documentclass[11pt]{article}
\usepackage[dvips]{graphicx}
\usepackage{enumitem}
\usepackage{amssymb}
\usepackage{color}
\usepackage{amsmath}
\usepackage{nicefrac}
\usepackage[left]{lineno}
\usepackage{hyperref}
\usepackage{adjustbox}
\usepackage{xspace}
\usepackage{multirow}
\usepackage{alphalph}
\usepackage[dvipsnames]{xcolor}

\definecolor{blu}{rgb}{0.,0.,1.}
\definecolor{red}{rgb}{1.,0.,0.}
\definecolor{burgundy}{rgb}{0.5, 0.0, 0.13}
\definecolor{crimsonred}{rgb}{0.6, 0.0, 0.0}
\definecolor{persianblue}{rgb}{0.11,0.22,0.73}
\definecolor{forestgreen}{rgb}{0.13,0.35,0.13}

\def\geant {\mbox{\textsc{Geant4}}\xspace}

\hypersetup{colorlinks, citecolor=crimsonred, linkcolor=persianblue, urlcolor=crimsonred}

\begin{document}
\centerline{\LARGE EUROPEAN ORGANIZATION FOR NUCLEAR RESEARCH}

% This is to be uncommented after CERN preprint ID is obtained
%
\vspace{10mm} {\flushright{
CERN-EP-2023-133 \\
10 July 2023\\
\vspace{4mm}
Revised version:\\30 August 2023\\
}}
\vspace{-30mm}

% This is for drafts
%
%\vspace{10mm}
%{\flushright{
%Draft 4\\
%21 June 2023\\
%}}
%\vspace{-15mm}

%%%%%%%%%%%%%%%%%%%%%
%
\vspace{40mm}

\begin{center}
\boldmath
{\bf {\LARGE\boldmath{Search for $K^+$ decays into the $\pi^+e^+e^-e^+e^-$ final state}}}
\unboldmath
\end{center}
\vspace{4mm}
\begin{center}
{\Large The NA62 Collaboration}\\
\end{center}

\begin{abstract}
The first search for ultra-rare $K^+$ decays into the $\pi^+e^+e^-e^+e^-$ final state is reported, using a dataset collected by the NA62 experiment at CERN in 2017--2018. An upper limit of \mbox{$1.4\times 10^{-8}$} at 90\% CL is obtained for the branching ratio of the $K^+\to\pi^+e^+e^-e^+e^-$ decay, predicted in the Standard Model to be $(7.2\pm0.7)\times 10^{-11}$. Upper limits at 90\% CL are obtained at the level of $10^{-9}$ for the branching ratios of two prompt decay chains involving pair-production of hidden-sector mediators: $K^+\to\pi^+aa$, $a\to e^+e^-$ and $K^+\to\pi^+S$, $S\to A^\prime A^\prime$, $A^\prime\to e^+e^-$.
\end{abstract}

\vspace{10mm}
\begin{center}
{\it Accepted for publication in Physics Letters B}
\end{center}

%\begin{linenumbers}

%%%%%%%%%%%%%%%%%%%%%%%%%%%%%%%%%%%%%%%

\newpage

\section*{Introduction}
%\vspace{-0.5mm}

Dark-sector models provide plausible dark-matter candidates, and represent a compelling new-physics direction to explore in its own right~\cite{be20,ag21}. So far, searches for the production of dark-sector mediators in meson decays have been focused on the production of a single particle, which is either invisible or decays into lepton or photon pairs. However, the parameter space of minimal dark sectors not yet excluded by experiment accommodates detectable pair-production of dark states~\cite{ho22,go23}. In the kaon sector, the process of particular interest is $K\to\pi XX$ followed by prompt $X\to e^+e^-$ decays, leading to characteristic multi-electron final states. Since this process has not been studied experimentally so far, ${\cal O}(10^{-6})$ sensitivity to its branching ratio is sufficient to improve existing constraints on dark-sector models.

The principal $K^+$ decay channel leading to the $\pi^+e^+e^-e^+e^-$ final state is the $K^+\to\pi^+\pi^0$ decay followed by the double Dalitz decay $\pi^0_{\rm DD}\to e^+e^-e^+e^-$. This decay chain, denoted $K_{2\pi {\rm DD}}$, with a measured branching ratio of $(6.9\pm0.3)\times 10^{-6}$~\cite{pdg}, is used for normalisation in this analysis. The non-resonant part of the  $K^+\to\pi^+e^+e^-e^+e^-$ decay, denoted $K_{\pi4e}$, is expected to occur via one-photon ($K^+\to\pi^+\gamma^*$) and two-photon ($K^+\to\pi^+\gamma^*\gamma^*$) exchange with an expected Standard Model (SM) branching ratio of $(7.2\pm0.7)\times 10^{-11}$~\cite{husek22}. The resulting final state is of two-fold interest in the context of dark sectors:
\vspace{-0.7mm}
\begin{itemize}
\item A short-lived QCD axion ($a$) decaying into an $e^+e^-$ pair is not completely ruled out by experiment~\cite{al18,al21,al23}, and provides a plausible explanation for the ``17~MeV anomaly'' in the mass spectra of the $e^+e^-$ pairs produced in the de-excitation of $^8$Be~\cite{x17-be8}, $^4$He~\cite{x17-he4} and $^{12}$C~\cite{x17-c12} nuclei. Assuming an axion mass of 17~MeV/$c^2$, lower bounds of ${\cal B}(K_L\to\pi^0 aa)>10^{-7}$ and ${\cal B}(K^+\to\pi^+aa)>2\times 10^{-8}$ (differing by the $K_L/K^+$ lifetime ratio) are predicted~\cite{ho22}, allowing a test of the QCD axion explanation for the 17~MeV anomaly.
\vspace{-0.7mm}
\item A scenario involving a dark scalar ($S$), and a dark photon ($A^\prime$) with masses satisfying the condition $m_S\ge 2m_{A^\prime}$ leads to a prompt cascade process $K^+\to\pi^+S$, $S\to A^\prime A^\prime$~\cite{ho22}. The case of prompt $A^\prime\to e^+e^-$ decay leads to the same final state as the $K_{\pi4e}$ decay.
%The case of the dark photon predominantly decaying promptly, $A^\prime\to e^+e^-$, has not been addressed by direct searches for dark-scalar production or beam-dump experiments.
\vspace{-0.7mm}
\end{itemize}

The NA62 experiment at CERN collected a dataset of $K^+$ decays into final states containing lepton pairs in 2017--2018. The first search, performed using this dataset, is reported here for the $K_{\pi4e}$ decay, and for pair-production of hidden-sector mediators in the prompt decay chains $K^+\to\pi^+aa$, $a\to e^+e^-$ (denoted $K_{\pi aa}$) and $K^+\to\pi^+S$, $S\to A^\prime A^\prime$, $A^\prime\to e^+e^-$ (denoted $K_{\pi S}$).

%\vspace{-1mm}

%%%%%%%%%%%%%%%%%%%%%%%%%%%%%%%%%%%%%%%%%%%%%%%%%

\section{Beam, detector and data sample}
\label{sec:detector}
%\vspace{-0.5mm}

\begin{figure}[t]
\begin{center}
\vspace{-2.9mm}
\resizebox{\textwidth}{!}{\includegraphics{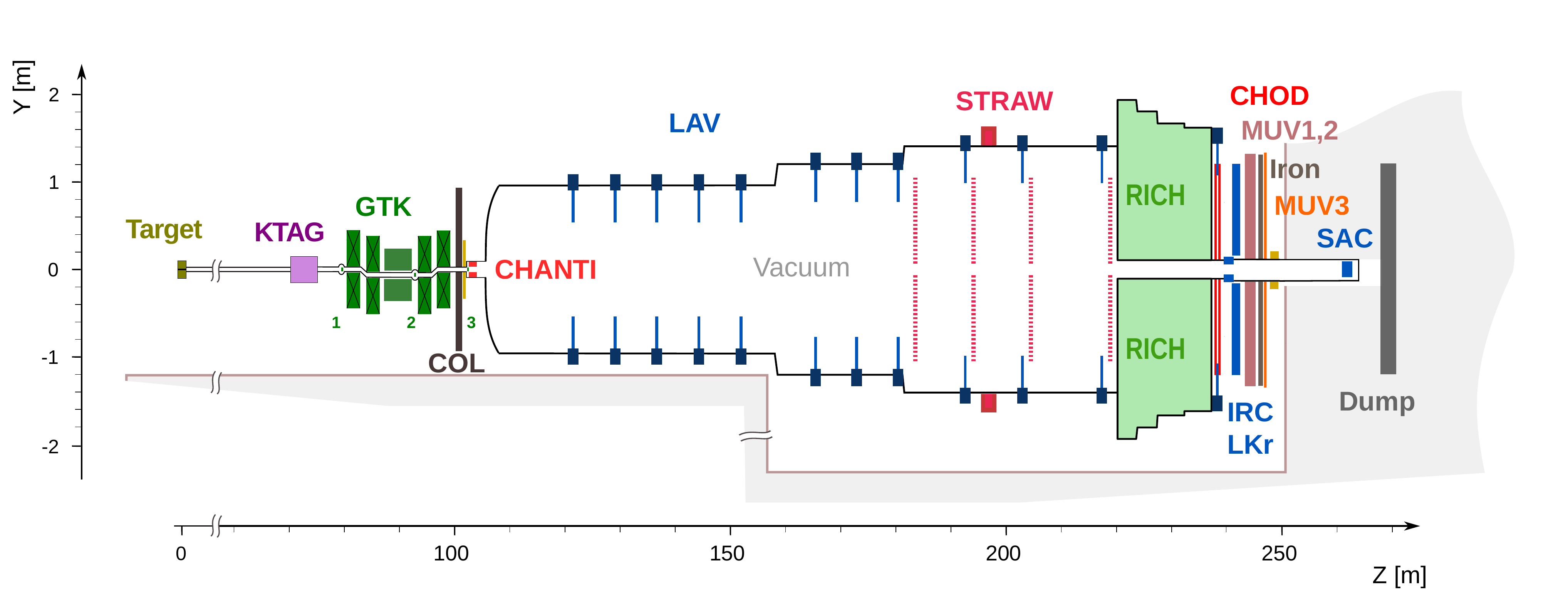}}
\put(-343,68){\scriptsize\color{OliveGreen}\rotatebox{90}{\textbf{\textsf{SCR}}}}
%\put(-325,121){\scriptsize\color{Sepia}\rotatebox{90}{\textbf{\textsf{COL}}}}
\put(-171.5,46){\scriptsize\color{red}{\textbf{\textsf{M}}}}
\put(-319.5,103){\tiny\color{YellowOrange}\rotatebox{90}{\textbf{\textsf{TRIM5}}}}
\end{center}
\vspace{-16.5mm}
\caption{Schematic side view of the NA62 beamline and detector used in 2018.}
\label{fig:detector}
\end{figure}

The layout of the NA62 beamline and detector~\cite{na62-detector} is shown schematically in Fig.~\ref{fig:detector}. An unseparated secondary beam of $\pi^+$ (70\%), protons (23\%) and $K^+$ (6\%) is created by directing 400~GeV/$c$ protons extracted from the CERN SPS onto a beryllium target in spills of 3~s effective duration. The central beam momentum is 75~GeV/$c$, with a momentum spread of 1\% (rms).

Beam kaons are tagged with a 70~ps time resolution by a differential Cherenkov counter (KTAG) that uses nitrogen gas at 1.75~bar pressure contained in a 5~m long vessel as the radiator. Beam particle positions, momenta and times
%(to a sub-100~ps resolution)
are measured by a silicon pixel spectrometer consisting of three stations (GTK1,2,3) and four dipole magnets forming an achromat; a toroidal muon sweeper (scraper, SCR) is installed between GTK1 and GTK2. A 1.2~m thick steel collimator (COL) with a $76\times40$~mm$^2$ central aperture and $1.7\times1.8$~m$^2$ outer dimensions is placed upstream of GTK3 to absorb hadrons from upstream $K^+$ decays; a variable-aperture collimator of $0.15\times0.15$~m$^2$ outer dimensions was used up to early 2018. A dipole magnet (TRIM5) providing a 90 MeV/c horizontal momentum kick is located in front of GTK3. Inelastic interactions of beam particles in GTK3 are detected by an array of scintillator hodoscopes (CHANTI). The beam is delivered into a vacuum tank evacuated to $10^{-6}$~mbar, which contains a 75~m long fiducial volume (FV) starting 2.6~m downstream of GTK3. The beam angular spread at the FV entrance is 0.11~mrad (rms) in both horizontal and vertical planes. Downstream of the FV, undecayed beam particles continue their path in vacuum.

Momenta of charged particles produced in $K^+$ decays in the FV are measured by a magnetic spectrometer (STRAW) located in the vacuum tank downstream of the FV. The spectrometer consists of four tracking chambers made of straw tubes, and a dipole magnet (M) located between the second and third chambers that provides a horizontal momentum kick of 270~MeV/$c$ in a direction opposite to that produced by TRIM5. The momentum resolution is $\sigma_p/p = (0.30\oplus 0.005\cdot p)\%$, with the momentum $p$ expressed in GeV/$c$.

% RICH threshold for pions quoted according to the detector paper.
% Another estimate: 90 * m_pi = 12.6 GeV/c.
%
A ring-imaging Cherenkov detector (RICH) consisting of a 17.5~m long vessel filled with neon at atmospheric pressure provides particle identification, charged particle time measurements with a typical resolution of 70~ps, and the trigger time. Two scintillator hodoscopes (CHOD), which include a matrix of tiles and two planes of slabs arranged in four quadrants located downstream of the RICH, provide trigger signals and time measurements with 200~ps precision.

A $27X_0$ thick quasi-homogeneous liquid-krypton (LKr) electromagnetic calorimeter is used for particle identification and photon detection. The calorimeter has an active volume of 7~m$^3$, segmented in the transverse direction into 13248 projective cells of $2\times 2$~cm$^2$ size, and provides an energy resolution $\sigma_E/E=(4.8/\sqrt{E}\oplus11/E\oplus0.9)\%$, with $E$ expressed in GeV. To achieve hermetic acceptance for photons emitted in $K^+$ decays in the FV at angles up to 50~mrad from the beam axis, the LKr calorimeter is complemented by annular lead glass detectors (LAV) installed in 12~positions inside and downstream of the vacuum tank, and two lead/scintillator sampling calorimeters (IRC, SAC) located close to the beam axis. An iron/scintillator sampling hadronic calorimeter formed of two modules (MUV1,2) and a muon detector consisting of 148~scintillator tiles located behind an 80~cm thick iron wall (MUV3) are used for particle identification.

%
% Data sample used:
% 17 [301k bursts per Twiki] + 18 [525k bursts per Twiki] = 826k bursts
%

The data sample analysed is obtained from $8.3\times 10^5$ SPS spills recorded in 2017--2018, with the typical beam intensity increasing over time from \mbox{$1.8\times 10^{12}$} to \mbox{$2.2\times 10^{12}$} protons per spill. The latter value corresponds to a 500~MHz mean beam particle rate at the FV entrance, and a 3.7~MHz mean $K^+$ decay rate in the FV. The main trigger line is designed for the measurement of the ultra-rare $K^+\to\pi^+\nu\bar\nu$ decay~\cite{pinn}. The analysis presented here
is based on dedicated electron multi-track ($e$MT) and multi-track (MT) trigger lines operating concurrently with the main trigger line~\cite{am19,na62-trigger}, typically downscaled by factors of 8 and 100, respectively. Both trigger lines involve a low-level stage and a high-level stage. The low-level (L0) hardware trigger is based on RICH signal multiplicity and coincidence of signals in two diagonally opposite CHOD quadrants, with an additional requirement of at least 20~GeV deposited in the LKr calorimeter in the $e$MT line. The high-level (L1) software trigger involves beam $K^+$ identification by the KTAG and reconstruction of a negatively-charged STRAW track.

Monte Carlo (MC) simulation of particle interactions with the detector and its response is performed using a software package based on the~\geant toolkit~\cite{geant4}. Accidental activity and trigger line response are included in the simulation.

%%%%%%%%%%%%%%%%%%%%%%%%%%%%%%%%%%%%%%%%%%%%%%%%%%%

\vspace{-1.8mm}
\section{Event selection}
\label{sec:selection}
\vspace{-1mm}

The $K_{2\pi {\rm DD}}$ events collected concurrently with signal candidates with the same trigger lines are used for normalisation. Signal and normalisation decay modes have the same set of particles in the final state, leading to a first-order cancellation of detector and trigger inefficiencies.

Five-track decays often produce final states with multiple tracks not fully contained in the geometric acceptance of the downstream detectors. The effect is particularly pronounced for the $K_{\pi 4e}$ and $K_{2\pi {\rm DD}}$ decays which typically produce soft, almost collinear electrons. For example, requiring a fully reconstructed five-track vertex leads to an acceptance of 1.2\% for the $K_{2\pi {\rm DD}}$ decay, and requiring track separation in the STRAW1 plane decreases the acceptance by another order of magnitude, which determines the overall acceptance level.

Only 11\% of the $K_{2\pi {\rm DD}}$ decays with a fully reconstructed five-track vertex have all tracks in the CHOD's geometric acceptance. To avoid an order-of-magnitude loss of acceptance, the analysis relies on the STRAW information only, and the STRAW tracks are not required to be in the geometric acceptance of any downstream detectors. The reduction in selectivity caused by the lack of information from the downstream detectors is mitigated by the exploitation
%compensated by the availability
of kinematic constraints on the five-track final state. Main selection criteria are described below.

\vspace{1.4mm}
\noindent Five-track vertices compatible with $K_{\pi4e}$ and $K_{2\pi\rm DD}$ decays are selected as follows:
\begin{itemize}
\vspace{-1mm}
\item Vertices are reconstructed by extrapolating STRAW tracks backward, taking into account the measured residual magnetic field in the vacuum tank. Exactly one five-track vertex should be present in the event within the FV. The longitudinal position of the vertex along the beam axis, $z_{\rm vtx}$, should be at least 5~m downstream of the start of the FV.
%to reduce the background from decays occurring upstream of GTK3 (which are misreconstructed due to the bending of particle trajectories in the TRIM5 magnet), the vertex should not be located in the first 5~meters of the FV.
%The distance between the vertex and the beam axis should not exceed 40~mm.
The total electric charge of the five tracks should be $q=+1$, their momenta should be in the range 5--45~GeV/$c$, and their trajectories through the STRAW chambers should be within the respective geometric acceptances. To suppress photon conversions and fake tracks, each pair of tracks should be separated by at least 15~mm in each STRAW chamber plane.
\vspace{-1mm}
\item The momentum excess defined as $\Delta p=p_{\rm vtx}-p_{\rm beam}$, where $p_{\rm vtx}$ is the total momentum of the five tracks and $p_{\rm beam}$ is the central beam momentum, should satisfy $|\Delta p|<2~{\rm GeV}/c$. The total transverse momentum of the tracks with respect to the beam axis should be $p_T<25~{\rm MeV}/c$. The quantity $p_{\rm beam}$ and the beam axis direction are monitored throughout the data taking, typically every few hours, with $K^+\to\pi^+\pi^+\pi^-$ decays.
\vspace{-1mm}
\item Track times are evaluated using the STRAW information, with a resolution of 6~ns. Each track is required to be within 20~ns of the trigger time.
\end{itemize}
\vspace{-1.2mm}
Particle identification is based on event kinematics, and is applied as follows:
\begin{itemize}
\vspace{-1mm}
\item The $K^+\to\pi^+e^+e^-e^+e^-$ final state involves three positively charged tracks (one $\pi^+$ and two $e^+$), and two negatively charged tracks (two $e^-$). Three assignments of the $\pi^+$ mass to one of the positively charged tracks are considered. In each mass assignment, the five-track mass, $m_{\pi 4e}$, the four-electron mass, $m_{4e}$, and the squared missing mass, $m_{\rm miss}^2=(P_K-P_\pi)^2$, are evaluated, where $P_K$ is the kaon four-momentum computed using the central beam momentum, and $P_\pi$ is the reconstructed $\pi^+$ four-momentum.
\vspace{-1mm}
\item The mass assignment corresponding to the minimal value of $|m_{\pi 4e}-m_K|$ is chosen, where $m_K$ is the $K^+$ mass, and it is required that $484~{\rm MeV}/c^2<m_{\pi4e}<504~{\rm MeV}/c^2$. The resolution of $m_{\pi4e}$ is 1.7~MeV/$c^2$. The probability of incorrectly assigning the mass is determined by simulation and found to be 4.5\% for $K_{\pi 4e}$ decays after the full selection, and ${\cal O}(10^{-4})$ for $K_{2\pi {\rm DD}}$ decays after the full selection which involves a $\pi^0$ mass constraint.
\end{itemize}
\noindent
The following condition is specific to $K_{2\pi\rm DD}$ selection:
\begin{itemize}
\item For the mass assignment chosen, it is required that $|m_{4e}-m_{\pi^0}|<10~{\rm MeV}/c^2$, where $m_{\pi^0}$ is the $\pi^0$ mass. The resolution of $m_{4e}$ is 0.9~MeV/$c^2$.
\end{itemize}
The following criteria are specific to $K_{\pi 4e}$ selection:
\begin{itemize}
\item To reject events consistent with a $K_{2\pi{\rm DD}}$ decay in any mass assignment, the requirements $|m_{4e}-m_{\pi^0}|>10~{\rm MeV}/c^2$, $m_{\rm miss}^2>0$ and $|m_{\rm miss}-m_{\pi^0}|>40~{\rm MeV}/c^2$ are applied for all three mass assignments. This reduces the $K_{\pi4e}$ acceptance by a factor of 0.74.
\item To suppress the $K^+\to\pi^+\pi^0_{\rm D}\pi^0_{\rm D}$ background, which predominantly enters by incorrect mass assignment (with a soft $e^+$ from a $\pi^0_{\rm D}\to\gamma e^+e^-$ decay considered as the $\pi^+$ candidate), the reconstructed $\pi^+$ momentum is required to be $p_\pi>10~{\rm GeV}/c$. This reduces the $K_{\pi4e}$ acceptance by a factor of 0.93.
\end{itemize}
Selection of the $K_{\pi aa}$ and $K_{\pi S}$ decay chains consists of the $K_{\pi 4e}$ selection with the following additional criteria:
\begin{itemize}
\item Two $X\to e^+e^-$ decay hypotheses are examined for the four $e^\pm$ candidates in the mass assignment chosen. In each hypothesis, the masses of the two $e^+e^-$ pairs, $m_{ee1}$ and $m_{ee2}$, are evaluated, and a discriminant is computed to quantify the consistency of the two mass values:
\begin{displaymath}
{\cal D}=(m_{ee1}-m_{ee2})^2 \, / \, (4.9\times 10^{-3}\cdot m_{ee})^2.
\end{displaymath}
Here $m_{ee} = (m_{ee1}+m_{ee2})/2$ is the reconstructed mass of the intermediate particle $X$ which is pair-produced, and the denominator represents the squared resolution of the quantity $m_{ee1}-m_{ee2}$, evaluated with simulation and found to be proportional to $m_{ee}$. The $X\to e^+e^-$ hypothesis with the smaller ${\cal D}$ value is considered, and it is required that ${\cal D}<9$ in this hypothesis.
\item For each $X$ mass hypothesis ($m_X$), it is required that $|m_{ee}-m_X|<0.02\cdot m_X$.
\end{itemize}

The signal region in the data is kept masked until the completion and validation of the background evaluation.

%%%%%%%%%%%%%%%%%%%%%%%%%%%%%

\boldmath
\section{Number of kaon decays in the fiducial volume}
\unboldmath
\label{sec:k2pidd}

\begin{figure}[t]
\begin{center}
\resizebox{0.5\textwidth}{!}{\includegraphics{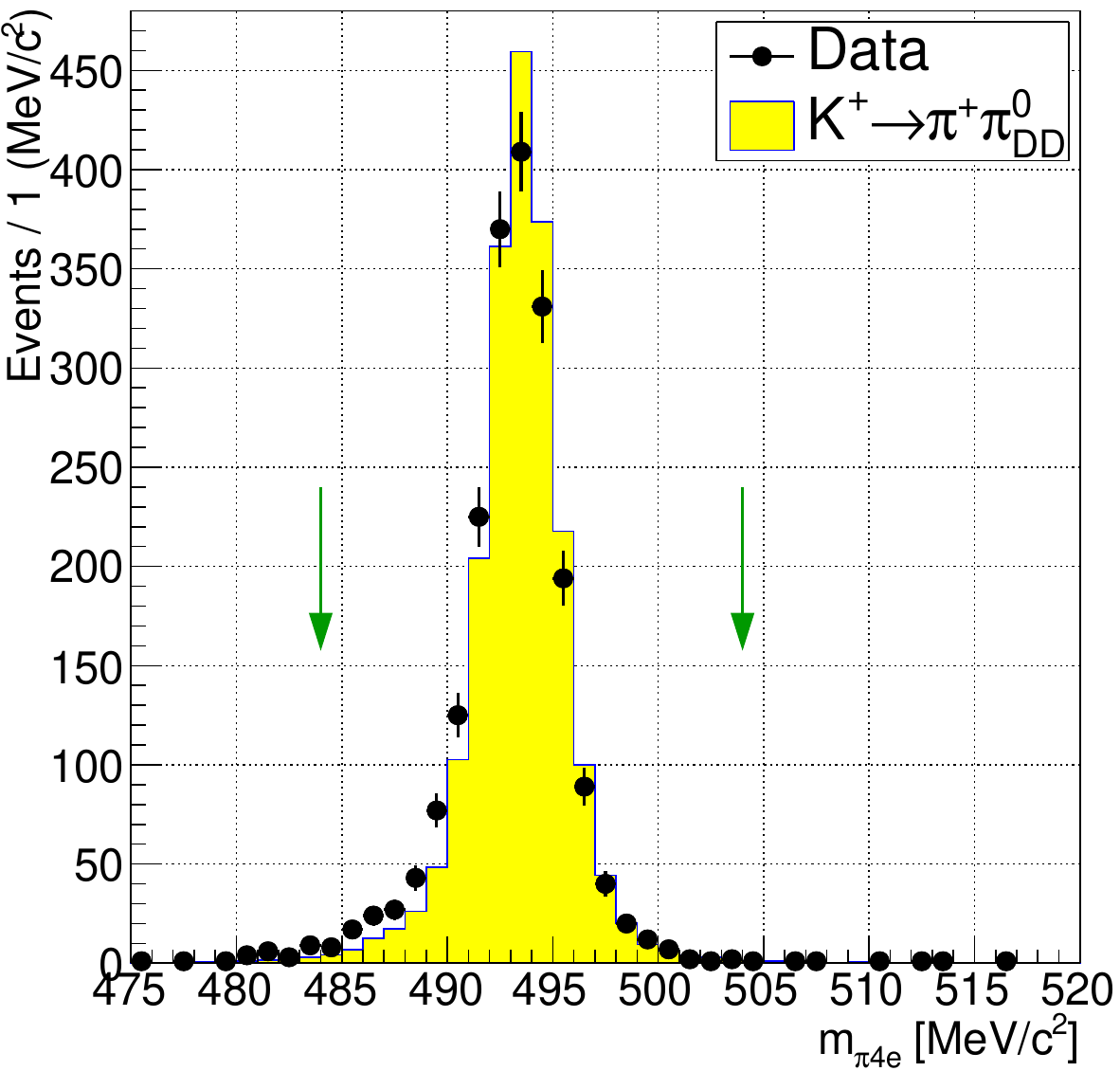}}%
\end{center}
\vspace{-15mm}
\caption{Reconstructed $m_{\pi4e}$ spectra for the data and a simulated $K_{2\pi \rm DD}$ sample obtained with the $K_{2\pi \rm DD}$ selection. The signal mass region is indicated by vertical arrows.}
\vspace{-2mm}
\label{fig:k2pidd}
\end{figure}

The reconstructed $m_{\pi 4e}$ distributions of $K_{2\pi{\rm DD}}$ candidates from both data and simulation are displayed in Fig.~\ref{fig:k2pidd}. The simulation reproduces the gaussian part of the $m_{\pi 4e}$ spectrum but does not fully describe the radiative tail; this is due to a soft-photon bremsstrahlung approximation~\cite{ka18} used to model the radiative corrections to the $\pi^0_{\rm DD}$ decay.

The number of $K^+$ decays in the FV is computed as
\begin{displaymath}
\vspace{-0.5mm}
N_K = \frac{f \cdot  N_{\rm DD}} {{\cal B}(K_{2\pi}) \cdot {\cal B}(\pi^0_{\rm DD}) \cdot A_{\rm DD}}
= (8.58 \pm 0.19_{\rm stat} \pm 0.07_{\rm MC} \pm 0.41_{\rm ext})\times 10^{11},
\vspace{-0.5mm}
\end{displaymath}
where $N_{\rm DD}=2023$ is the number of $K_{2\pi{\rm DD}}$ candidates in the data;
${\cal B}(K_{2\pi})=(20.67\pm0.08)\times 10^{-2}$
and ${\cal B}(\pi^0_{\rm DD})=(3.34\pm0.16)\times 10^{-5}$ are the $K^+\to\pi^+\pi^0$ and $\pi^0_{\rm DD}$ decay branching ratios~\cite{pdg}; $A_{\rm DD}=(3.41 \pm 0.02_{\rm stat}\pm 0.02_{\rm syst})\times 10^{-4}$ is the acceptance of the $K_{2\pi{\rm DD}}$ selection evaluated with simulation (the systematic error is evaluated by variation of the $m_{\pi4e}$ selection conditions); $f=0.9987$ is the purity of the $K_{2\pi{\rm DD}}$ sample evaluated with simulation  and accounting for the $K^+\to\pi^0_{\rm DD}\ell^+\nu$ background ($\ell=e,\mu$). The three uncertainties quoted in $N_K$ arise from the limited size of the data sample, the total uncertainty in the acceptance $A_{\rm DD}$, and the external uncertainties in the measured values of ${\cal B}(K_{2\pi})$ and ${\cal B}(\pi^0_{\rm DD})$.

The acceptance $A_{\rm DD}$ includes the effects of accidental activity and trigger efficiencies. The efficiencies of the CHOD and LKr trigger conditions for the $K_{2\pi\rm DD}$ sample are found with simulation to be 74\% and 81\% respectively, and are relatively low due to the absence of the CHOD and LKr geometric acceptance requirements for the tracks. The efficiencies of the RICH, KTAG and STRAW trigger conditions exceed 99\%.

%%%%%%%%%%%%%%%%%%%%%%%%%%%%%%

\section{Background estimation}
\label{sec:bkg}

Background from five-track and seven-track $K^+$ decays is evaluated with simulation; all $K^+$ decays with branching ratios exceeding $10^{-8}$ are considered. Background from coincidences of pairs of three-track $K^+$ decays in time (with one track not reconstructed) is evaluated with dedicated simulation of pairs of decays occurring within 25~ns; coincidences of all types of three-track $K^+$ decays with products of branching ratios exceeding $2\times 10^{-6}$ are considered.
%using the branching ratios and combinatorial factors.
The probability of two decays occurring within this 25~ns time window is dependent upon the instantaneous beam intensity, and is accounted for by normalising the total estimated background to the data in the momentum excess region \mbox{$\Delta p>20~{\rm GeV}/c$}.

Contributions from all possible triple coincidences of one or more $K^+\to\pi^+\pi^+\pi^-$ decays with $K^+\to\mu^+\nu$ decays and beam halo particles, and contributions from coincidences of three-track decays with inelastic interactions of beam particles in GTK3, are found to be much smaller than those from coincidences of pairs of three-track decays, and are neglected.

\begin{figure}[p]
\begin{center}
\resizebox{0.5\textwidth}{!}{\includegraphics{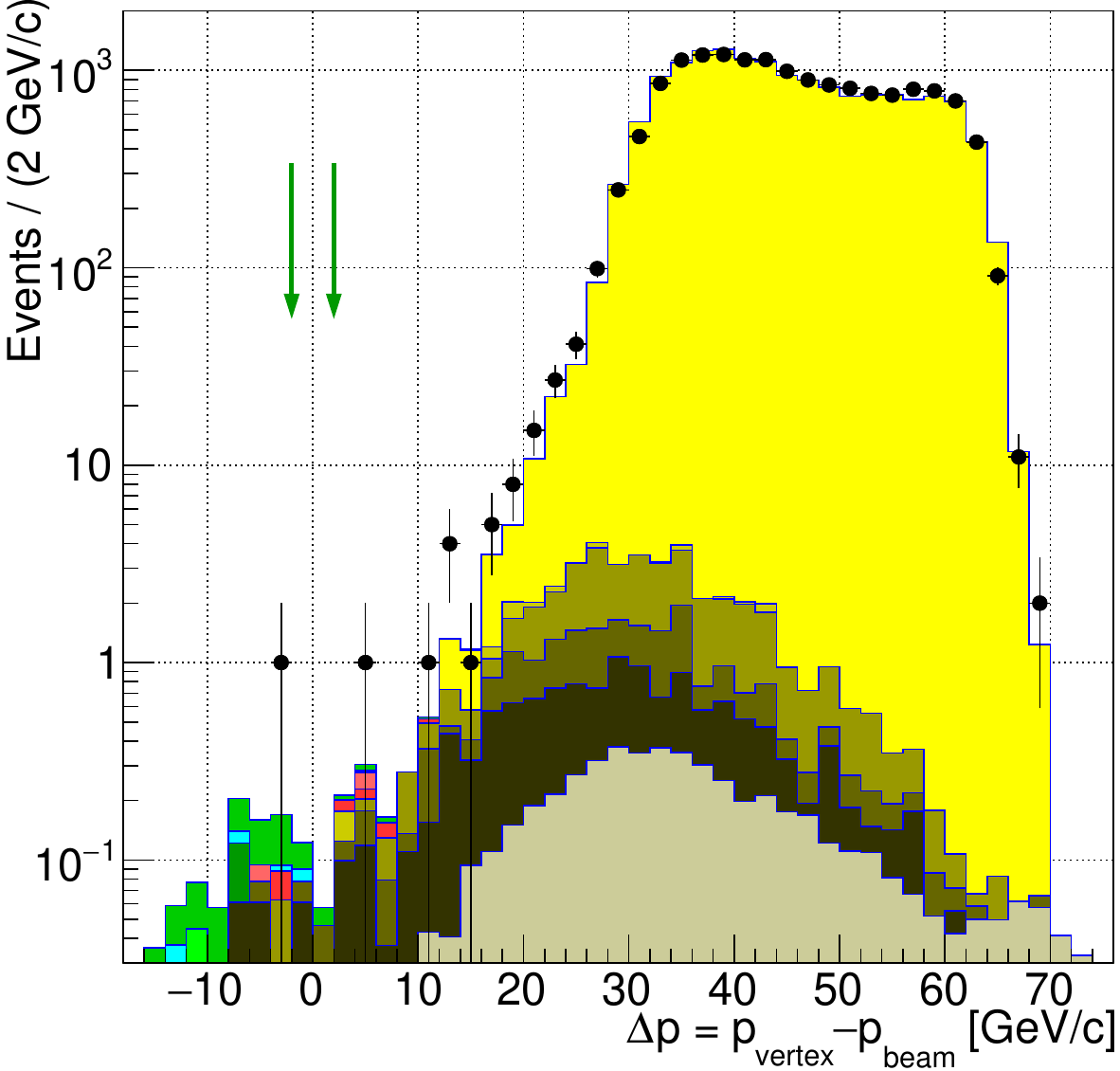}}%
\resizebox{0.5\textwidth}{!}{\includegraphics{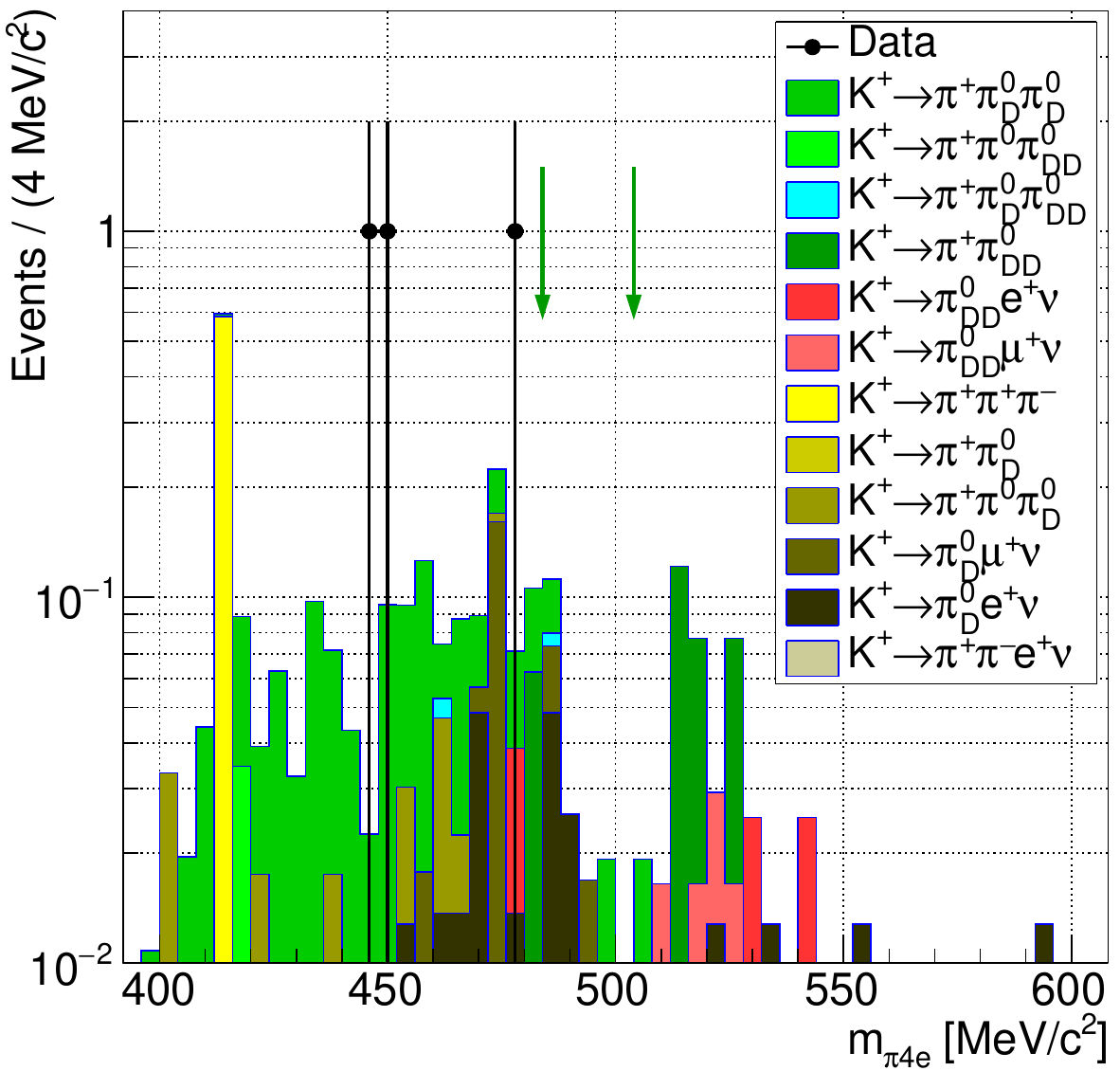}}%
\end{center}
\vspace{-15mm}
\caption{Reconstructed momentum excess ($\Delta p$, left) and five-track mass ($m_{\pi 4e}$, right) spectra for data and simulated samples after the $K_{\pi 4e}$ selection obtained by removing the $\Delta p$ condition (left panel) or the $m_{\pi 4e}$ condition (right panel).
%The $m_{\pi 4e}$ condition is applied to obtain the $\Delta p$ spectrum, and the $\Delta p$ condition is applied to obtain the $m_{\pi 4e}$ spectrum.
Contributions from three-track $K^+$ decays (the last six entries in the legend) are due to coincidences with a $K^+\to\pi^+\pi^+\pi^-$ decay. The sum of these contributions is normalised to the data in the region $\Delta p>20~{\rm GeV}/c$. The signal region is shown with pairs of vertical arrows in both plots.}
\label{fig:kpi4e}
\end{figure}

The reconstructed momentum excess ($\Delta p$) and five-track mass ($m_{\pi4e}$) spectra for data and simulated events after the $K_{\pi 4e}$ selection are shown in Fig.~\ref{fig:kpi4e}. The region of large $\Delta p$ values is dominated by backgrounds from coincidences of three-track decays. The integrals of the data and simulated spectra agree in the region $\Delta p>20~{\rm GeV}/c$ by construction. The region $\Delta p<0$ is dominated by backgrounds from single five-track decays which, with the exception of the $K_{2\pi\rm DD}$ decay, lead to photons and neutrinos in the final state.

The background model is validated against the data in the control region $\Delta p<-2~{\rm GeV}/c$ using the standard $K_{\pi4e}$ selection (Fig.~\ref{fig:kpi4e}, left) and a loose $K_{\pi4e}$ selection with the condition \mbox{$p_\pi>10~{\rm GeV}/c$} removed. The results are shown in Table~\ref{tab:bkg}. The loose selection leads to a sharp increase in the $K^+\to\pi^+\pi^0_{\rm D}\pi^0_{\rm D}$ background in the control region due to the soft momentum spectrum of the positrons from $\pi^0_{\rm D}$ decays misidentified as $\pi^+$ (Section~\ref{sec:selection}).

%%%%%%%%%%%%%%%%%%%%%%%%%%%%%%%%%%%%%%%%

\begin{table}[p]
\caption{Backgrounds to the $K_{\pi 4e}$ decay, their branching ratios (or products of branching ratios for coincidences of two decays), and estimated backgrounds with their statistical uncertainties in the control ($\Delta p<-2~{\rm GeV}/c$) and signal ($|\Delta p|<2~{\rm GeV}/c$) momentum excess regions shown in Fig.~\ref{fig:kpi4e}~(left). Estimated backgrounds in the control region for a loose $K_{\pi4e}$ selection, with the $p_\pi>10~{\rm GeV}/c$ condition removed, are also shown. The $K^+\to\pi^+\pi^0_{\rm D}e^+e^-$ decay, as well as the $K^+\to\pi^+\pi^-e^+\nu$ decay in coincidence with a $K^+\to\pi^+\pi^+\pi^-$ decay, and coincidences of any of the two decays, $K^+\to\pi^+\pi^+\pi^-$ and $K^+\to\pi^+\pi^0_{\rm D}$, lead to negligible backgrounds in the control and signal regions, and are not listed.}
\begin{center}
\vspace{-8mm}
\begin{tabular}{lcccc}
\hline
Source & Branching ratio & Control region & Control region, & Signal region \\
& (or their product) & & loose selection & \\
\hline
\multicolumn{4}{l}{Single five-track and seven-track decays}\\
\hline
$K_{2\pi{\rm DD}}$ & $6.9\times10^{-6}$ & $0.06\pm0.06$ & $0.06\pm0.06$ & -- \\
$K^+\to\pi^+\pi^0_{\rm D}\pi^0_{\rm D}$ & $2.4\times10^{-6}$ & $0.30\pm0.06$ &
$2.47\pm0.16$ & $0.04\pm0.02$\\
$K^+\to\pi^0_{\rm DD}e^+\nu$ & $1.7\times10^{-6}$ & $0.10\pm0.05$ & $0.10\pm0.05$ & -- \\
$K^+\to\pi^+\pi^0\pi^0_{\rm DD}$ & $1.2\times10^{-6}$ & $0.03\pm0.03$ & $0.03\pm0.03$ & -- \\
$K^+\to\pi^0_{\rm DD}\mu^+\nu$ & $1.1\times10^{-6}$ & $0.02\pm0.02$ & $0.03\pm0.02$ & -- \\
$K^+\to\pi^+\pi^0_{\rm D}\pi^0_{\rm DD}$ & $1.4\times10^{-8}$ & $0.05\pm0.02$ &
$0.10\pm0.02$ & $0.01\pm0.01$ \\
\hline
\multicolumn{4}{l}{Coincidences of three-track decays with a $K^+\to\pi^+\pi^+\pi^-$ decay}\\
\hline
$K^+\to\pi^0_{\rm D}e^+\nu$ & $3.3\times10^{-5}$ & $0.15\pm0.07$ & $0.15\pm0.07$ & $0.08\pm0.05$ \\
$K^+\to\pi^+\pi^0\pi^0_{\rm D}$ & $2.3\times10^{-5}$ & $0.03\pm0.03$ & $0.08\pm0.05$ & -- \\
$K^+\to\pi^0_{\rm D}\mu^+\nu$ & $2.2\times10^{-5}$ & $0.03\pm0.02$ & $0.04\pm0.02$ & $0.05\pm0.02$ \\
\hline
Total && $0.77\pm0.13$ & $3.06\pm0.21$ & $0.18\pm0.06$ \\
\hline
Data && 1 & 4 & 0 \\
\hline
\end{tabular}
\end{center}
\label{tab:bkg}
\end{table}

%%%%%%%%%%%%%%%%

The background model is further validated using the $m_{\pi4e}$ spectrum outside the signal region. Using the standard $K_{\pi4e}$ selection, three data events are observed with an expected background of $2.83\pm0.64_{\rm stat}$ (Fig.~\ref{fig:kpi4e}, right); using the loose $K_{\pi4e}$ selection, five data events are observed with an expected background of $3.99\pm0.74_{\rm stat}$. In all cases, the data are in agreement with simulation within statistical fluctuations, and the largest background comes from the $K^+\to\pi^+\pi^0_{\rm D}\pi^0_{\rm D}$ decay.

The background contributions in the signal region after the $K_{\pi4e}$ selection are summarised in Table~\ref{tab:bkg}. The estimate of the total background, validated by agreement of the data with simulation in the control regions, is
\begin{displaymath}
N_B = 0.18\pm0.06_{\rm stat} \pm 0.13_{\rm syst} = 0.18 \pm 0.14,
\end{displaymath}
where a 100\% relative systematic uncertainty is assigned to the contribution from coincidences of three-track decays, based on the level of agreement of data and simulation in the momentum excess region $2~{\rm GeV}/c<\Delta p<20~{\rm GeV}/c$. 

The requirement that the two reconstructed $e^+e^-$ masses ($m_{ee1}$, $m_{ee2}$) are consistent with each other, applied in the selection of the $K_{\pi aa}$ and $K_{\pi S}$ decay chains, leads to a reduction of the background with respect to the $K_{\pi4e}$ selection. The probability of the reconstructed $m_{ee1}$ and $m_{ee2}$ values being consistent is found to be 2\% in simulated samples of $K^+\to\pi^+\pi^0_{\rm D}\pi^0_{\rm D}$ and $K^+\to\pi^+\pi^0_{\rm DD}$ decays. Furthermore, the probability of $|m_{ee}-m_X|<0.02\cdot m_X$ for the above samples does not exceed 10\% in any $m_X$ hypothesis considered (where $X=a,A'$) as established with simulations. The background to the $K_{\pi aa}$ and $K_{\pi S}$ processes in each $m_X$ hypothesis is estimated as
\begin{displaymath}
N_B^\prime = (0.4\pm0.4)\times 10^{-3},
\end{displaymath}
where a 100\% relative uncertainty is conservatively assigned.

%%%%%%%%%%%%%%%%

\section{Search for the signal decays}

The $K_{\pi4e}$ decay is modelled according to the SM description~\cite{husek22}. The non-resonant part of the decay amplitude is considered, including the dominant one-photon exchange diagram ($K^+\to\pi^+\gamma^*$), the two-photon exchange diagram ($K^+\to\pi^+\gamma^*\gamma^*$), and their interference. The contribution from the $\pi^0$ pole in the two-photon exchange is neglected; it does not interfere with the other amplitudes, peaks at $m_{4e}\approx m_{\pi^0}$, and accounts for 3\% of the total decay rate in the region $|m_{4e}-m_{\pi^0}|>10~{\rm MeV}/c^2$ used in this analysis. The acceptance of the $K_{\pi 4e}$ selection for the $K_{\pi4e}$ signal is found to be $A_{\pi 4e}=(1.85\pm0.06_{\rm stat})\times 10^{-4}$.

The prompt decay chain $K^+\to\pi^+aa$, $a\to e^+e^-$, allowed kinematically for axion masses in the range $2m_e \le m_a \le (m_K-m_\pi)/2$, is simulated with a uniform $K^+\to\pi^+aa$ phase-space distribution~\cite{ho22} and isotropic $a\to e^+e^-$ decays, for 33~equally-spaced $m_a$ hypotheses in the range 10--170~MeV/$c^2$. The acceptance of the $K_{\pi aa}$ selection reaches a maximum of $7.1\times 10^{-3}$ for $m_a=155~{\rm MeV}/c^2$.

The prompt decay chain $K^+\to\pi^+S$, $S\to A^\prime A^\prime$, $A^\prime\to e^+e^-$, allowed kinematically in the mass region $2m_e \le m_{A^\prime} \le m_S/2 \le (m_K-m_\pi)/2$, is simulated as a series of isotropic decays for $m_{A^\prime}$ hypotheses in the range 10--170~MeV/$c^2$ and $m_S$ hypotheses in the range 20--340~MeV/$c^2$, with a 5~MeV/$c^2$ step in both $m_{A^\prime}$ and $m_S$. The region $120~{\rm MeV}/c^2<m_S<165~{\rm MeV}/c^2$ is excluded from the search as the signal acceptance is suppressed in this region by the selection conditions that reduce the $K_{2\pi\rm DD}$ background. The acceptance of the $K_{\pi S}$ selection reaches a maximum of $1.1\times 10^{-2}$ at the kinematic boundary for $m_{A^\prime}=150~{\rm MeV}/c^2$ and $m_S=300~{\rm MeV}/c^2$.

The $K_{\pi aa}$ and $K_{\pi S}$ decays typically produce energetic, well-separated electrons in the final state, which leads to significantly larger acceptances than for $K_{\pi 4e}$ and $K_{2\pi\rm DD}$ decays.

%%%%%%%%%%%%%%%%%%%%%%%%%%%%%%%

\section{Results}

No data events are observed in the signal region after unmasking. In this case, upper limits at 90\% CL on the signal branching ratios computed using the CL$_{\rm S}$ method~\cite{re02} do not depend on the expected background and are equal to $2.3\,/(A\,N_K)$, where $A$ denotes the signal acceptance. The uncertainties in $N_K$ and signal acceptances are taken into account in the CL$_{\rm S}$ procedure.
An upper limit of ${\cal B}(K_{\pi4e}) < 1.4\times 10^{-8}$ is established at 90\%~CL for the branching ratio of the non-resonant part of the $K_{\pi4e}$ decay, which is a factor 200 larger than the SM expectation~\cite{husek22}.

\begin{figure}[tb]
\begin{center}
\resizebox{0.5\textwidth}{!}{\includegraphics{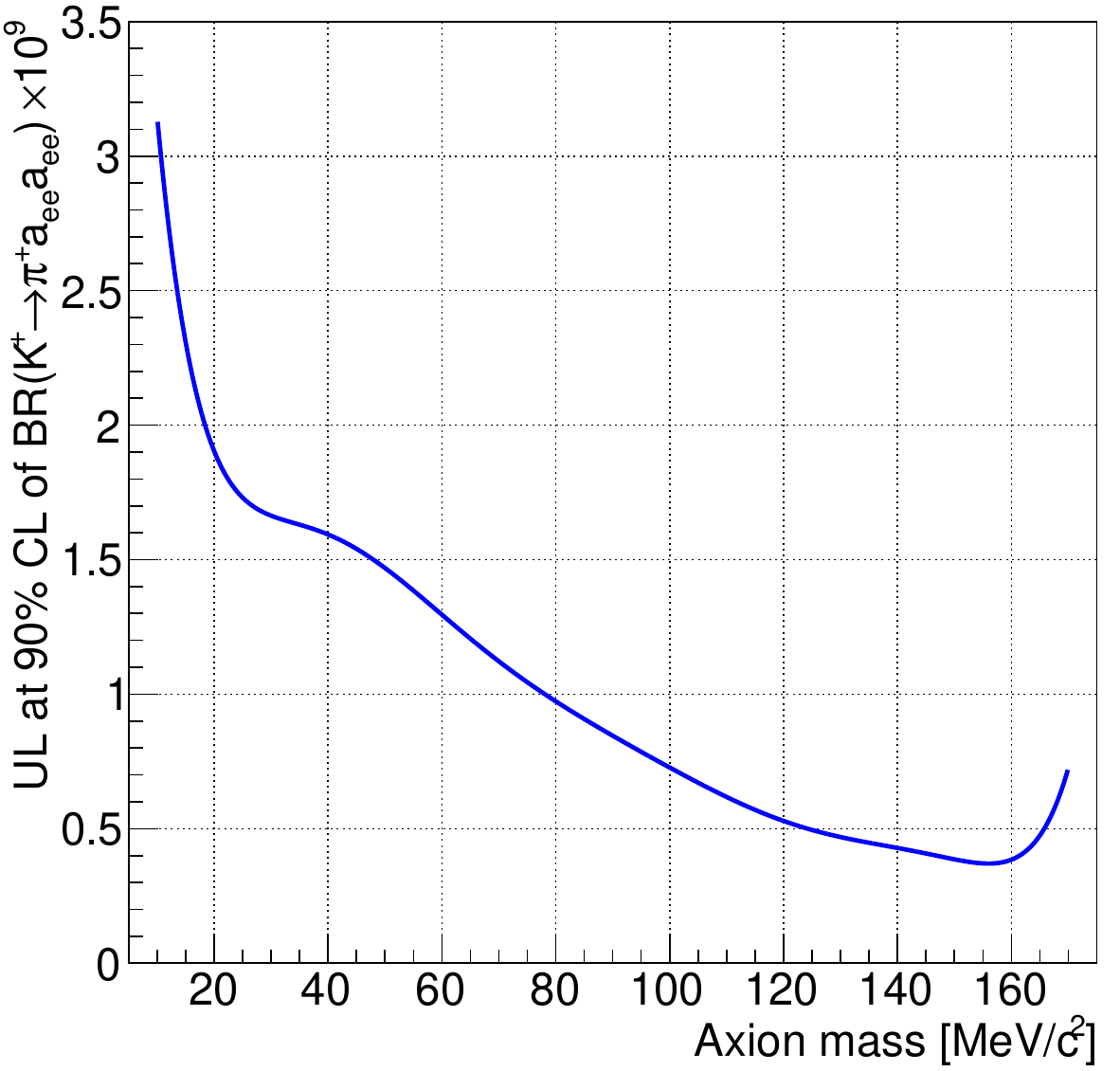}}%
\resizebox{0.5\textwidth}{!}{\includegraphics{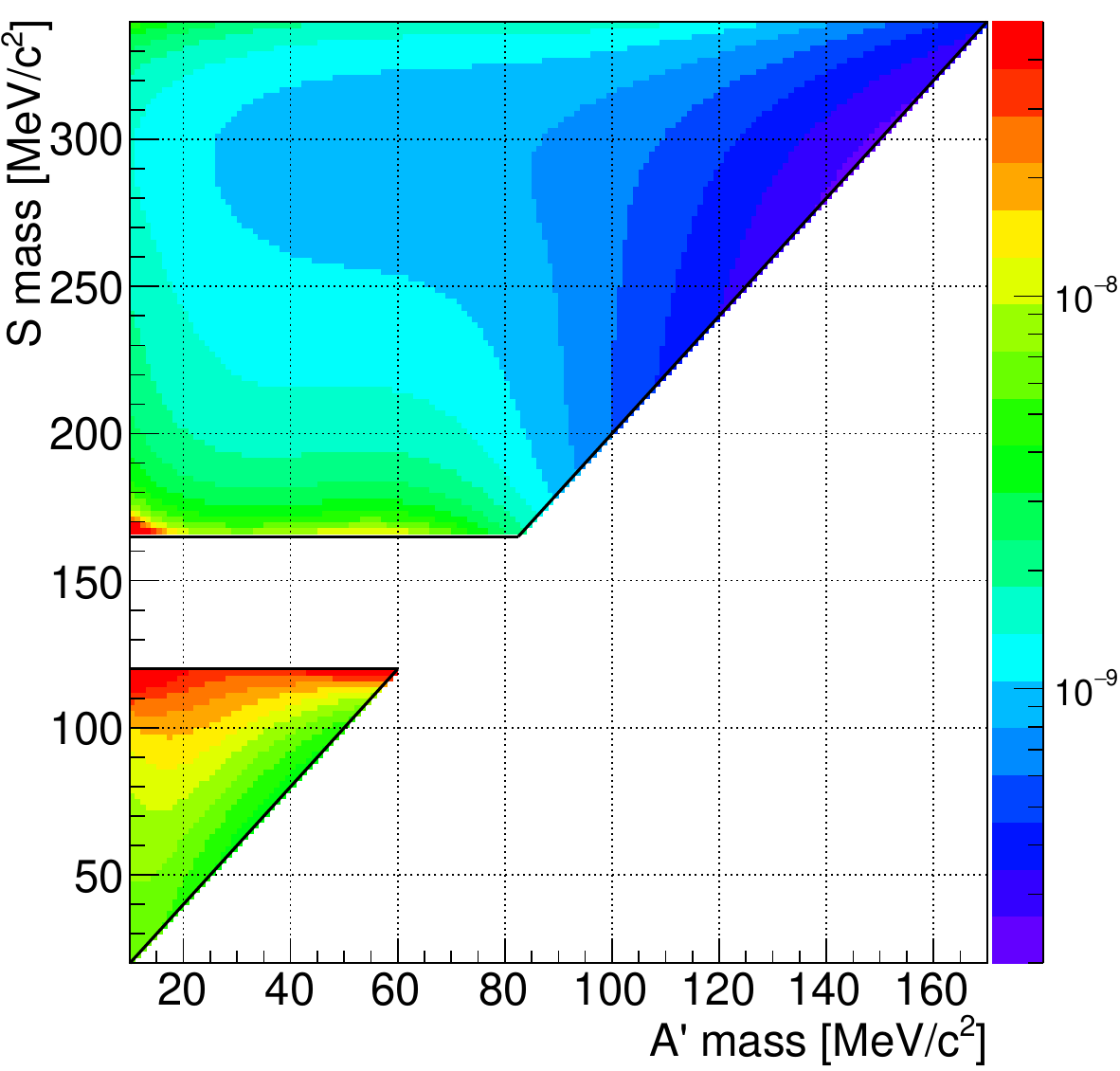}}%
\end{center}
\vspace{-15mm}
\caption{Left: upper limit at 90\%~CL of the branching ratio of the prompt decay chain $K^+\to\pi^+aa$, $a\to e^+e^-$ as a function of the assumed axion mass. Right: upper limit at 90\%~CL of the branching ratio of the prompt decay chain $K^+\to\pi^+S$, $S\to A^\prime A^\prime$, $A^\prime\to e^+e^-$ as a function of the assumed dark-photon and dark-scalar masses. Discrete sets of points obtained with a step size of 5~MeV/$c^2$ are approximated by polynomial functions in each case.}
\vspace{-2mm}
\label{fig:limits}
\end{figure}

Upper limits obtained at 90\%~CL for the branching ratios of the $K_{\pi aa}$ and $K_{\pi S}$ decay chains are shown as functions of the assumed masses of the dark-sector mediators in Fig.~\ref{fig:limits}. For the product ${\cal B}(K^+\to\pi^+aa)\times\left[{\cal B}(a\to e^+e^-)\right]^2$, the most stringent upper limit of $3.7\times 10^{-10}$ is obtained for $m_a=155~{\rm MeV}/c^2$. The upper limit of $2.1\times 10^{-9}$ obtained for the above quantity for $m_a=17~{\rm MeV}/c^2$ excludes the QCD axion explanation of the ``17~MeV anomaly'' in the mass spectra of the $e^+e^-$ pairs produced in nuclear de-excitation~\cite{x17-be8, x17-he4, x17-c12}, according to the decay rate estimate reported in Ref.~\cite{ho22}. For the product ${\cal B}(K^+\to\pi^+S)\times{\cal B}(S\to A^\prime A^\prime)\times\left[{\cal B}(A^\prime\to e^+e^-)\right]^2$, the most stringent upper limit of $2.5\times 10^{-10}$ is obtained for $m_{A^\prime}=150~{\rm MeV}/c^2$ and $m_S=300~{\rm MeV}/c^2$.

%%%%%%%%%%%%%%%%%%%%%%%%%%%%%%%%%%%%%%%%%%%

\section*{Summary}

The first search for ultra-rare $K^+$ decays into the $\pi^+e^+e^-e^+e^-$ final state is reported, using a dataset collected by the NA62 experiment at CERN in 2017--2018. An upper limit of $1.4\times 10^{-8}$ at 90\%~CL is obtained for the branching ratio of the $K^+\to\pi^+e^+e^-e^+e^-$ decay, with the expected SM branching ratio of $(7.2\pm0.7)\times 10^{-11}$. Upper limits at 90\% CL are obtained at the level of $10^{-9}$ for the branching ratios of two prompt decay chains involving pair-production of hidden-sector mediators: $K^+\to\pi^+aa$, $a\to e^+e^-$ and $K^+\to\pi^+S$, $S\to A^\prime A^\prime$, $A^\prime\to e^+e^-$. The QCD axion is excluded as a possible explanation of the ``17~MeV anomaly'' according to the decay rate estimate reported in Ref.~\cite{ho22}.

%%%%%%%%%%%%%%%%%%%%%%%%%%%%%%%%%%%%%%%%%%%%

\section*{Acknowledgements}

We are grateful to Matheus Hostert and Maxim Pospelov for suggesting this study, and for the discussions of the phenomenology of pair-production of dark states.
It is a pleasure to express our appreciation to the staff of the CERN laboratory and the technical
staff of the participating laboratories and universities for their efforts in the operation of the
experiment and data processing.

The cost of the experiment and its auxiliary systems was supported by the funding agencies of 
the Collaboration Institutes. We are particularly indebted to: 
F.R.S.-FNRS (Fonds de la Recherche Scientifique - FNRS), under Grants No. 4.4512.10, 1.B.258.20, Belgium;
CECI (Consortium des Equipements de Calcul Intensif), funded by the Fonds de la Recherche Scientifique de Belgique (F.R.S.-FNRS) under Grant No. 2.5020.11 and by the Walloon Region, Belgium;
%BMES (Ministry of Education, Youth and Science), Bulgaria;
NSERC (Natural Sciences and Engineering Research Council), funding SAPPJ-2018-0017,  Canada;
%NRC (National Research Council) contribution to TRIUMF, Canada;
MEYS (Ministry of Education, Youth and Sports) funding LM 2018104, Czech Republic;
BMBF (Bundesministerium f\"{u}r Bildung und Forschung) contracts 05H12UM5, 05H15UMCNA and 05H18UMCNA, Germany;
INFN  (Istituto Nazionale di Fisica Nucleare),  Italy;
MIUR (Ministero dell'Istruzione, dell'Universit\`a e della Ricerca),  Italy;
CONACyT  (Consejo Nacional de Ciencia y Tecnolog\'{i}a),  Mexico;
IFA (Institute of Atomic Physics) Romanian 
% ended in 2019
CERN-RO No. 1/16.03.2016 
% 2020 and 2021
%CERN-RO Nr. 10/10.03.2020
% 2022-2024
%CERN-RO Nr. 06/03.01.2022
and Nucleus Programme PN 19 06 01 04,  Romania;
% remove for opt C
%INR-RAS (Institute for Nuclear Research of the Russian Academy of Sciences), Moscow, Russia; 
%JINR (Joint Institute for Nuclear Research), Dubna, Russia; 
%NRC (National Research Center)  ``Kurchatov Institute'' and MESRF (Ministry of Education and Science of the Russian Federation), Russia; 
MESRS  (Ministry of Education, Science, Research and Sport), Slovakia; 
CERN (European Organization for Nuclear Research), Switzerland; 
STFC (Science and Technology Facilities Council), United Kingdom;
NSF (National Science Foundation) Award Numbers 1506088 and 1806430,  U.S.A.;
ERC (European Research Council)  ``UniversaLepto'' advanced grant 268062, ``KaonLepton'' starting grant 336581, Europe.

Individuals have received support from:
Charles University Research Center (UNCE/SCI/013), Czech Republic;
Ministero dell'Istruzione, dell'Universit\`a e della Ricerca (MIUR  ``Futuro in ricerca 2012''  grant RBFR12JF2Z, Project GAP), Italy;
% terminnated
% Russian Foundation for Basic Research  (RFBR grants 18-32-00072, 18-32-00245), Russia; 
% remove for opt C
%Russian Science Foundation (RSF 19-72-10096), Russia;
the Royal Society  (grants UF100308, UF0758946), United Kingdom;
STFC (Rutherford fellowships ST/J00412X/1, ST/M005798/1), United Kingdom;
ERC (grants 268062,  336581 and  starting grant 802836 ``AxScale'');
EU Horizon 2020 (Marie Sk\l{}odowska-Curie grants 701386, 754496, 842407, 893101, 101023808).
% remove for opt C
%The data used in this paper were collected before February 2022.

%\end{linenumbers}

%%%%%%%%%%%%%%%%%%%%%%%%%%%%%%%%%%%%%%%%%%%%

%\newpage

\newpage

\newcommand{\orcimg}{\raisebox{-0.3\height}{\includegraphics[height=\fontcharht\font`A]{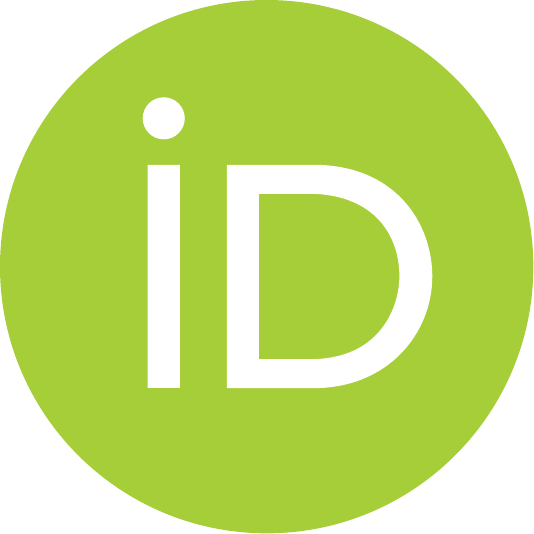}}}
\newcommand{\orcid}[1]{\href{https://orcid.org/#1}{\orcimg}}

\centerline{\bf The NA62 Collaboration} 
\vspace{0.5cm}
%
%%%%%%%%%%%%%%%%%%%%%%%%%%%%%%%%%
%

\begin{raggedright}
\noindent
%%%%%%%
{\bf Universit\'e Catholique de Louvain, Louvain-La-Neuve, Belgium}\\
 E.~Cortina Gil\orcid{0000-0001-9627-699X},
 A.~Kleimenova$\,${\footnotemark[1]}\orcid{0000-0002-9129-4985},
 E.~Minucci$\,${\footnotemark[2]}\orcid{0000-0002-3972-6824},
 S.~Padolski\orcid{0000-0002-6795-7670},
 P.~Petrov, 
 A.~Shaikhiev$\,${\footnotemark[3]}\orcid{0000-0003-2921-8743},
 R.~Volpe$\,${\footnotemark[4]}\orcid{0000-0003-1782-2978}
\vspace{0.5cm}

%%%%%%%
{\bf TRIUMF, Vancouver, British Columbia, Canada}\\
 T.~Numao\orcid{0000-0001-5232-6190},
 Y.~Petrov\orcid{0000-0003-2643-8740},
 B.~Velghe\orcid{0000-0002-0797-8381},
 V. W. S.~Wong\orcid{0000-0001-5975-8164}
\vspace{0.5cm}

%%%%%%%
{\bf University of British Columbia, Vancouver, British Columbia, Canada}\\
 D.~Bryman$\,${\footnotemark[5]}\orcid{0000-0002-9691-0775},
 J.~Fu
\vspace{0.5cm}

%%%%%%%
{\bf Charles University, Prague, Czech Republic}\\
 Z.~Hives\orcid{0000-0002-5025-993X},
 T.~Husek\orcid{0000-0002-7208-9150},
 J.~Jerhot$\,${\footnotemark[6]}\orcid{0000-0002-3236-1471},
 K.~Kampf\orcid{0000-0003-1096-667X},
 M.~Zamkovsky$\,${\footnotemark[7]}\orcid{0000-0002-5067-4789}
\vspace{0.5cm}

%%%%%%%
{\bf Aix Marseille University, CNRS/IN2P3, CPPM, Marseille, France}\\
 B.~De Martino\orcid{0000-0003-2028-9326},
 M.~Perrin-Terrin\orcid{0000-0002-3568-1956}
\vspace{0.5cm}

%%%%%%%
{\bf Institut f\"ur Physik and PRISMA Cluster of Excellence, Universit\"at Mainz, Mainz, Germany}\\
 A.T.~Akmete\orcid{0000-0002-5580-5477},
 R.~Aliberti$\,${\footnotemark[8]}\orcid{0000-0003-3500-4012},
 G.~Khoriauli$\,${\footnotemark[9]}\orcid{0000-0002-6353-8452},
 J.~Kunze,
 D.~Lomidze$\,${\footnotemark[10]}\orcid{0000-0003-3936-6942}, 
 L.~Peruzzo\orcid{0000-0002-4752-6160},
 M.~Vormstein,
 R.~Wanke\orcid{0000-0002-3636-360X}
\vspace{0.5cm}

%%%%%%%
{\bf Dipartimento di Fisica e Scienze della Terra dell'Universit\`a e INFN, Sezione di Ferrara, Ferrara, Italy}\\
 P.~Dalpiaz,
 M.~Fiorini\orcid{0000-0001-6559-2084},
 A.~Mazzolari\orcid{0000-0003-0804-6778},
 I.~Neri\orcid{0000-0002-9669-1058},
 A.~Norton$\,${\footnotemark[11]}\orcid{0000-0001-5959-5879}, 
 F.~Petrucci\orcid{0000-0002-7220-6919},
 M.~Soldani\orcid{0000-0003-4902-943X},
 H.~Wahl$\,${\footnotemark[12]}\orcid{0000-0003-0354-2465}
\vspace{0.5cm}

%%%%%%%
{\bf INFN, Sezione di Ferrara, Ferrara, Italy}\\
 L.~Bandiera\orcid{0000-0002-5537-9674},
 A.~Cotta Ramusino\orcid{0000-0003-1727-2478},
 A.~Gianoli\orcid{0000-0002-2456-8667},
 M.~Romagnoni\orcid{0000-0002-2775-6903},
 A.~Sytov\orcid{0000-0001-8789-2440}
\vspace{0.5cm}

%%%%%%%
{\bf Dipartimento di Fisica e Astronomia dell'Universit\`a e INFN, Sezione di Firenze, Sesto Fiorentino, Italy}\\
 E.~Iacopini\orcid{0000-0002-5605-2497},
 G.~Latino\orcid{0000-0002-4098-3502},
 M.~Lenti\orcid{0000-0002-2765-3955},
 P.~Lo Chiatto\orcid{0000-0002-4177-557X},
 I.~Panichi\orcid{0000-0001-7749-7914},
 A.~Parenti\orcid{0000-0002-6132-5680}
\vspace{0.5cm}

%%%%%%%
{\bf INFN, Sezione di Firenze, Sesto Fiorentino, Italy}\\
 A.~Bizzeti$\,${\footnotemark[13]}\orcid{0000-0001-5729-5530},
 F.~Bucci\orcid{0000-0003-1726-3838}
\vspace{0.5cm}

%%%%%%%
{\bf Laboratori Nazionali di Frascati, Frascati, Italy}\\
 A.~Antonelli\orcid{0000-0001-7671-7890},
 G.~Georgiev$\,${\footnotemark[14]}\orcid{0000-0001-6884-3942},
 V.~Kozhuharov$\,${\footnotemark[14]}\orcid{0000-0002-0669-7799},
 G.~Lanfranchi\orcid{0000-0002-9467-8001},
 S.~Martellotti\orcid{0000-0002-4363-7816}, 
 M.~Moulson\orcid{0000-0002-3951-4389},
 T.~Spadaro\orcid{0000-0002-7101-2389},
 G.~Tinti\orcid{0000-0003-1364-844X}
\vspace{0.5cm}

%%%%%%%
{\bf Dipartimento di Fisica ``Ettore Pancini'' e INFN, Sezione di Napoli, Napoli, Italy}\\
 F.~Ambrosino\orcid{0000-0001-5577-1820},
 T.~Capussela,
 M.~Corvino\orcid{0000-0002-2401-412X},
 M.~D'Errico\orcid{0000-0001-5326-1106},
 D.~Di Filippo\orcid{0000-0003-1567-6786}, 
 R.~Fiorenza$\,${\footnotemark[15]}\orcid{0000-0003-4965-7073},
 R.~Giordano\orcid{0000-0002-5496-7247},
 P.~Massarotti\orcid{0000-0002-9335-9690},
 M.~Mirra\orcid{0000-0002-1190-2961},
 M.~Napolitano\orcid{0000-0003-1074-9552}, 
 I.~Rosa\orcid{0009-0002-7564-182},
 G.~Saracino\orcid{0000-0002-0714-5777}
\vspace{0.5cm}

%%%%%%%
{\bf Dipartimento di Fisica e Geologia dell'Universit\`a e INFN, Sezione di Perugia, Perugia, Italy}\\
 G.~Anzivino\orcid{0000-0002-5967-0952},
 F.~Brizioli$\,${\footnotemark[7]}\orcid{0000-0002-2047-441X},
 E.~Imbergamo,
 R.~Lollini\orcid{0000-0003-3898-7464},
 R.~Piandani$\,${\footnotemark[16]}\orcid{0000-0003-2226-8924},
 C.~Santoni\orcid{0000-0001-7023-7116}
\vspace{0.5cm}

%%%%%%%
{\bf INFN, Sezione di Perugia, Perugia, Italy}\\
 M.~Barbanera\orcid{0000-0002-3616-3341},
 P.~Cenci\orcid{0000-0001-6149-2676},
 B.~Checcucci\orcid{0000-0002-6464-1099},
 P.~Lubrano\orcid{0000-0003-0221-4806},
 M.~Lupi$\,${\footnotemark[17]}\orcid{0000-0001-9770-6197}, 
 M.~Pepe\orcid{0000-0001-5624-4010},
 M.~Piccini\orcid{0000-0001-8659-4409}
\vspace{0.5cm}

%%%%%%%
{\bf Dipartimento di Fisica dell'Universit\`a e INFN, Sezione di Pisa, Pisa, Italy}\\
 F.~Costantini\orcid{0000-0002-2974-0067},
 L.~Di Lella$\,${\footnotemark[12]}\orcid{0000-0003-3697-1098},
 N.~Doble$\,${\footnotemark[12]}\orcid{0000-0002-0174-5608},
 M.~Giorgi\orcid{0000-0001-9571-6260},
 S.~Giudici\orcid{0000-0003-3423-7981}, 
 G.~Lamanna\orcid{0000-0001-7452-8498},
 E.~Lari\orcid{0000-0003-3303-0524},
 E.~Pedreschi\orcid{0000-0001-7631-3933},
 M.~Sozzi\orcid{0000-0002-2923-1465}
\vspace{0.5cm}

%%%%%%%
{\bf INFN, Sezione di Pisa, Pisa, Italy}\\
 C.~Cerri,
 R.~Fantechi\orcid{0000-0002-6243-5726},
 L.~Pontisso$\,${\footnotemark[18]}\orcid{0000-0001-7137-5254},
 F.~Spinella\orcid{0000-0002-9607-7920}
\vspace{0.5cm}

%%%%%%%
{\bf Scuola Normale Superiore e INFN, Sezione di Pisa, Pisa, Italy}\\
 I.~Mannelli\orcid{0000-0003-0445-7422}
\vspace{0.5cm}

%%%%%%%
{\bf Dipartimento di Fisica, Sapienza Universit\`a di Roma e INFN, Sezione di Roma I, Roma, Italy}\\
 G.~D'Agostini\orcid{0000-0002-6245-875X},
 M.~Raggi\orcid{0000-0002-7448-9481}
\vspace{0.5cm}

%%%%%%%
{\bf INFN, Sezione di Roma I, Roma, Italy}\\
 A.~Biagioni\orcid{0000-0001-5820-1209},
 P.~Cretaro\orcid{0000-0002-2229-149X},
 O.~Frezza\orcid{0000-0001-8277-1877},
 E.~Leonardi\orcid{0000-0001-8728-7582},
 A.~Lonardo\orcid{0000-0002-5909-6508}, 
 M.~Turisini\orcid{0000-0002-5422-1891},
 P.~Valente\orcid{0000-0002-5413-0068},
 P.~Vicini\orcid{0000-0002-4379-4563}
\vspace{0.5cm}

%%%%%%%
{\bf INFN, Sezione di Roma Tor Vergata, Roma, Italy}\\
 R.~Ammendola\orcid{0000-0003-4501-3289},
 V.~Bonaiuto$\,${\footnotemark[19]}\orcid{0000-0002-2328-4793},
 A.~Fucci,
 A.~Salamon\orcid{0000-0002-8438-8983},
 F.~Sargeni$\,${\footnotemark[20]}\orcid{0000-0002-0131-236X}
\vspace{0.5cm}

%%%%%%%
{\bf Dipartimento di Fisica dell'Universit\`a e INFN, Sezione di Torino, Torino, Italy}\\
 R.~Arcidiacono$\,${\footnotemark[21]}\orcid{0000-0001-5904-142X},
 B.~Bloch-Devaux\orcid{0000-0002-2463-1232},
 M.~Boretto$\,${\footnotemark[7]}\orcid{0000-0001-5012-4480},
 E.~Menichetti\orcid{0000-0001-7143-8200},
 E.~Migliore\orcid{0000-0002-2271-5192},
 D.~Soldi\orcid{0000-0001-9059-4831}
\vspace{0.5cm}

%%%%%%%
{\bf INFN, Sezione di Torino, Torino, Italy}\\
 C.~Biino\orcid{0000-0002-1397-7246},
 A.~Filippi\orcid{0000-0003-4715-8748},
 F.~Marchetto\orcid{0000-0002-5623-8494}
\vspace{0.5cm}

%%%%%%%
{\bf Instituto de F\'isica, Universidad Aut\'onoma de San Luis Potos\'i, San Luis Potos\'i, Mexico}\\
 A.~Briano Olvera\orcid{0000-0001-6121-3905},
 J.~Engelfried\orcid{0000-0001-5478-0602},
 N.~Estrada-Tristan$\,${\footnotemark[22]}\orcid{0000-0003-2977-9380},
 M. A.~Reyes Santos$\,${\footnotemark[22]}\orcid{0000-0003-1347-2579}
\vspace{0.5cm}

%%%%%%%
{\bf Horia Hulubei National Institute for R\&D in Physics and Nuclear Engineering, Bucharest-Magurele, Romania}\\
 P.~Boboc\orcid{0000-0001-5532-4887},
 A. M.~Bragadireanu,
 S. A.~Ghinescu\orcid{0000-0003-3716-9857},
 O. E.~Hutanu
\vspace{0.5cm}

%%%%%%%
{\bf Faculty of Mathematics, Physics and Informatics, Comenius University, Bratislava, Slovakia}\\
 L.~Bician$\,${\footnotemark[23]}\orcid{0000-0001-9318-0116},
 T.~Blazek\orcid{0000-0002-2645-0283},
 V.~Cerny\orcid{0000-0003-1998-3441},
 Z.~Kucerova$\,${\footnotemark[7]}\orcid{0000-0001-8906-3902}
\vspace{0.5cm}

%%%%%%%
{\bf CERN, European Organization for Nuclear Research, Geneva, Switzerland}\\
 J.~Bernhard\orcid{0000-0001-9256-971X},
 A.~Ceccucci\orcid{0000-0002-9506-866X},
 M.~Ceoletta\orcid{0000-0002-2532-0217},
 H.~Danielsson\orcid{0000-0002-1016-5576},
 N.~De Simone$\,${\footnotemark[24]}, 
 F.~Duval,
 B.~D\"obrich$\,${\footnotemark[25]}\orcid{0000-0002-6008-8601},
 L.~Federici\orcid{0000-0002-3401-9522},
 E.~Gamberini\orcid{0000-0002-6040-4985},
 L.~Gatignon$\,${\footnotemark[3]}\orcid{0000-0001-6439-2945}, 
 R.~Guida,
 F.~Hahn$\,$\renewcommand{\thefootnote}{\fnsymbol{footnote}}\footnotemark[2]\renewcommand{\thefootnote}{\arabic{footnote}},
 E.~B.~Holzer\orcid{0000-0003-2622-6844},
 B.~Jenninger,
 M.~Koval$\,${\footnotemark[23]}\orcid{0000-0002-6027-317X}, 
 P.~Laycock$\,${\footnotemark[26]}\orcid{0000-0002-8572-5339},
 G.~Lehmann Miotto\orcid{0000-0001-9045-7853},
 P.~Lichard\orcid{0000-0003-2223-9373},
 A.~Mapelli\orcid{0000-0002-4128-1019},
 R.~Marchevski$\,${\footnotemark[1]}\orcid{0000-0003-3410-0918}, 
 K.~Massri\orcid{0000-0001-7533-6295},
 M.~Noy,
 V.~Palladino\orcid{0000-0002-9786-9620},
 J.~Pinzino$\,${\footnotemark[27]}\orcid{0000-0002-7418-0636},
 V.~Ryjov, 
 S.~Schuchmann\orcid{0000-0002-8088-4226},
 S.~Venditti
\vspace{0.5cm}

%%%%%%%
{\bf School of Physics and Astronomy, University of Birmingham, Birmingham, United Kingdom}\\
 T.~Bache\orcid{0000-0003-4520-830X},
 M. B.~Brunetti$\,${\footnotemark[28]}\orcid{0000-0003-1639-3577},
 V.~Duk$\,${\footnotemark[4]}\orcid{0000-0001-6440-0087},
 V.~Fascianelli$\,${\footnotemark[29]},
 J. R.~Fry\orcid{0000-0002-3680-361X}, 
 F.~Gonnella\orcid{0000-0003-0885-1654},
 E.~Goudzovski$\,$\renewcommand{\thefootnote}{\fnsymbol{footnote}}\footnotemark[1]\renewcommand{\thefootnote}{\arabic{footnote}}\orcid{0000-0001-9398-4237},
 J.~Henshaw\orcid{0000-0001-7059-421X},
 L.~Iacobuzio,
 C.~Kenworthy\orcid{0009-0002-8815-0048}, 
 C.~Lazzeroni\orcid{0000-0003-4074-4787},
 N.~Lurkin$\,${\footnotemark[6]}\orcid{0000-0002-9440-5927},
 F.~Newson,
 C.~Parkinson\orcid{0000-0003-0344-7361},
 A.~Romano\orcid{0000-0003-1779-9122}, 
 J.~Sanders\orcid{0000-0003-1014-094X},
 A.~Sergi$\,${\footnotemark[30]}\orcid{0000-0001-9495-6115},
 A.~Sturgess\orcid{0000-0002-8104-5571},
 J.~Swallow$\,${\footnotemark[7]}\orcid{0000-0002-1521-0911},
 A.~Tomczak\orcid{0000-0001-5635-3567}
\vspace{0.5cm}

%%%%%%%
{\bf School of Physics, University of Bristol, Bristol, United Kingdom}\\
 H.~Heath\orcid{0000-0001-6576-9740},
 R.~Page,
 S.~Trilov\orcid{0000-0003-0267-6402}
\vspace{0.5cm}

%%%%%%%
{\bf School of Physics and Astronomy, University of Glasgow, Glasgow, United Kingdom}\\
 B.~Angelucci,
 D.~Britton\orcid{0000-0001-9998-4342},
 C.~Graham\orcid{0000-0001-9121-460X},
 D.~Protopopescu\orcid{0000-0002-3964-3930}
\vspace{0.5cm}

%%%%%%%
{\bf Faculty of Science and Technology, University of Lancaster, Lancaster, United Kingdom}\\
 J.~Carmignani$\,${\footnotemark[31]}\orcid{0000-0002-1705-1061},
 J. B.~Dainton,
 R. W. L.~Jones\orcid{0000-0002-6427-3513},
 G.~Ruggiero$\,${\footnotemark[32]}\orcid{0000-0001-6605-4739}
\vspace{0.5cm}

%%%%%%%
{\bf School of Physical Sciences, University of Liverpool, Liverpool, United Kingdom}\\
 L.~Fulton,
 D.~Hutchcroft\orcid{0000-0002-4174-6509},
 E.~Maurice$\,${\footnotemark[33]}\orcid{0000-0002-7366-4364},
 B.~Wrona\orcid{0000-0002-1555-0262}
\vspace{0.5cm}

%%%%%%%
{\bf Physics and Astronomy Department, George Mason University, Fairfax, Virginia, USA}\\
 A.~Conovaloff,
 P.~Cooper,
 D.~Coward$\,${\footnotemark[34]}\orcid{0000-0001-7588-1779},
 P.~Rubin\orcid{0000-0001-6678-4985}
\vspace{0.5cm}

%%%%%%%
{\bf Authors affiliated with an Institute or an international laboratory covered by a cooperation agreement with CERN}\\
 A.~Baeva,
 D.~Baigarashev$\,${\footnotemark[35]}\orcid{0000-0001-6101-317X},
 D.~Emelyanov,
 T.~Enik\orcid{0000-0002-2761-9730},
 V.~Falaleev$\,${\footnotemark[4]}\orcid{0000-0003-3150-2196}, 
 S.~Fedotov,
 K.~Gorshanov\orcid{0000-0001-7912-5962},
 E.~Gushchin\orcid{0000-0001-8857-1665},
 V.~Kekelidze\orcid{0000-0001-8122-5065},
 D.~Kereibay, 
 S.~Kholodenko$\,${\footnotemark[27]}\orcid{0000-0002-0260-6570},
 A.~Khotyantsev,
 A.~Korotkova,
 Y.~Kudenko\orcid{0000-0003-3204-9426},
 V.~Kurochka, 
 V.~Kurshetsov\orcid{0000-0003-0174-7336},
 L.~Litov$\,${\footnotemark[14]}\orcid{0000-0002-8511-6883},
 D.~Madigozhin\orcid{0000-0001-8524-3455},
 M.~Medvedeva,
 A.~Mefodev, 
 M.~Misheva$\,${\footnotemark[36]},
 N.~Molokanova,
 S.~Movchan,
 V.~Obraztsov\orcid{0000-0002-0994-3641},
 A.~Okhotnikov\orcid{0000-0003-1404-3522}, 
 A.~Ostankov$\,$\renewcommand{\thefootnote}{\fnsymbol{footnote}}\footnotemark[2]\renewcommand{\thefootnote}{\arabic{footnote}},
 I.~Polenkevich,
 Yu.~Potrebenikov\orcid{0000-0003-1437-4129},
 A.~Sadovskiy\orcid{0000-0002-4448-6845},
 V.~Semenov$\,$\renewcommand{\thefootnote}{\fnsymbol{footnote}}\footnotemark[2]\renewcommand{\thefootnote}{\arabic{footnote}}, 
 S.~Shkarovskiy,
 V.~Sugonyaev\orcid{0000-0003-4449-9993},
 O.~Yushchenko\orcid{0000-0003-4236-5115},
 A.~Zinchenko$\,$\renewcommand{\thefootnote}{\fnsymbol{footnote}}\footnotemark[2]\renewcommand{\thefootnote}{\arabic{footnote}}
\vspace{0.5cm}

\end{raggedright}

%
%%%%%%%%%%%%%%%%%%%%%%%%%%%%%%%%%
%

\setcounter{footnote}{0}
\newlength{\basefootnotesep}
\setlength{\basefootnotesep}{\footnotesep}

\renewcommand{\thefootnote}{\fnsymbol{footnote}}
\noindent
$^{\footnotemark[1]}${Corresponding author: E.~Goudzovski, email: evgueni.goudzovski@cern.ch}\\
$^{\footnotemark[2]}${Deceased}\\
\renewcommand{\thefootnote}{\arabic{footnote}}
$^{1}${Present address: Ecole Polytechnique F\'ed\'erale Lausanne, CH-1015 Lausanne, Switzerland} \\
$^{2}${Present address: Syracuse University, Syracuse, NY 13244, USA} \\
$^{3}${Present address: Faculty of Science and Technology, University of Lancaster, Lancaster, LA1 4YW, UK} \\
$^{4}${Present address: INFN, Sezione di Perugia, I-06100 Perugia, Italy} \\
$^{5}${Also at TRIUMF, Vancouver, British Columbia, V6T 2A3, Canada} \\
$^{6}${Present address: Universit\'e Catholique de Louvain, B-1348 Louvain-La-Neuve, Belgium} \\
$^{7}${Present address: CERN, European Organization for Nuclear Research, CH-1211 Geneva 23, Switzerland} \\
$^{8}${Present address: Institut f\"ur Kernphysik and Helmholtz Institute Mainz, Universit\"at Mainz, Mainz, D-55099, Germany} \\
$^{9}${Present address: Universit\"at W\"urzburg, D-97070 W\"urzburg, Germany} \\
$^{10}${Present address: European XFEL GmbH, D-22869 Schenefeld, Germany} \\
$^{11}${Present address: School of Physics and Astronomy, University of Glasgow, Glasgow, G12 8QQ, UK} \\
$^{12}${Present address: Institut f\"ur Physik and PRISMA Cluster of Excellence, Universit\"at Mainz, D-55099 Mainz, Germany} \\
$^{13}${Also at Dipartimento di Scienze Fisiche, Informatiche e Matematiche, Universit\`a di Modena e Reggio Emilia, I-41125 Modena, Italy} \\
$^{14}${Also at Faculty of Physics, University of Sofia, BG-1164 Sofia, Bulgaria} \\
$^{15}${Present address: Scuola Superiore Meridionale e INFN, Sezione di Napoli, I-80138 Napoli, Italy} \\
$^{16}${Present address: Instituto de F\'isica, Universidad Aut\'onoma de San Luis Potos\'i, 78240 San Luis Potos\'i, Mexico} \\
$^{17}${Present address: Institut am Fachbereich Informatik und Mathematik, Goethe Universit\"at, D-60323 Frankfurt am Main, Germany} \\
$^{18}${Present address: INFN, Sezione di Roma I, I-00185 Roma, Italy} \\
$^{19}${Also at Department of Industrial Engineering, University of Roma Tor Vergata, I-00173 Roma, Italy} \\
$^{20}${Also at Department of Electronic Engineering, University of Roma Tor Vergata, I-00173 Roma, Italy} \\
$^{21}${Also at Universit\`a degli Studi del Piemonte Orientale, I-13100 Vercelli, Italy} \\
$^{22}${Also at Universidad de Guanajuato, 36000 Guanajuato, Mexico} \\
$^{23}${Present address: Charles University, 116 36 Prague 1, Czech Republic} \\
$^{24}${Present address: DESY, D-15738 Zeuthen, Germany} \\
$^{25}${Present address: Max-Planck-Institut f\"ur Physik (Werner-Heisenberg-Institut), M\"unchen, D-80805, Germany} \\
$^{26}${Present address: Brookhaven National Laboratory, Upton, NY 11973, USA} \\
$^{27}${Present address: INFN, Sezione di Pisa, I-56100 Pisa, Italy} \\
$^{28}${Present address: Department of Physics, University of Warwick, Coventry, CV4 7AL, UK} \\
$^{29}${Present address: Center for theoretical neuroscience, Columbia University, New York, NY 10027, USA} \\
$^{30}${Present address: Dipartimento di Fisica dell'Universit\`a e INFN, Sezione di Genova, I-16146 Genova, Italy} \\
$^{31}${Present address: School of Physical Sciences, University of Liverpool, Liverpool, L69 7ZE, UK} \\
$^{32}${Present address: Dipartimento di Fisica e Astronomia dell'Universit\`a e INFN, Sezione di Firenze, I-50019 Sesto Fiorentino, Italy} \\
$^{33}${Present address: Laboratoire Leprince Ringuet, F-91120 Palaiseau, France} \\
$^{34}${Also at SLAC National Accelerator Laboratory, Stanford University, Menlo Park, CA 94025, USA} \\
$^{35}${Also at L.N. Gumilyov Eurasian National University, 010000 Nur-Sultan, Kazakhstan} \\
$^{36}${Present address: Institute of Nuclear Research and Nuclear Energy of Bulgarian Academy of Science (INRNE-BAS), BG-1784 Sofia, Bulgaria} \\

\clearpage

\end{document}